# The anomalous Hall Effect in a Magnetically Extended Topological Insulator Heterostructure

Nan Liu[†,‡], Xuefan Niu[†,‡], Yuxin Liu[†,‡], Qinghua Zhang[*,†,‡], Lin Gu[†,‡], Jing Teng[*,†,‡], and Yongqing Li[†,‡,⊥]

[†]Beijing National Laboratory for Condensed Matter Physics, Institute of Physics, Chinese Academy of Sciences, Beijing 100190, China

[‡]School of Physical Sciences, University of Chinese Academy of Sciences, Beijing 100049, China

[⊥] Beijing Key Laboratory for Nanomaterials and Nanodevices, Beijing 100190, China

ABSTRACT: Constructing heterostructures of a topological insulator (TI) with an *undoped* magnetic insulator (MI) is a clean and versatile approach to break the time-reversible symmetry in the TI surface states. Despite a lot of efforts, the strength of interfacial magnetic proximity effect (MPE) is still too weak to achieve the quantum anomalous Hall effect and many other topological quantum phenomena. Recently, a new approach based on intercalation of atomic layers of MI, referred to as magnetic extension, was proposed to realize strong MPE [2D Mater. 4, 025082(2017)]. Motivated by this proposal, here, we study a magnetic extension system prepared by molecular beam epitaxy growth of MnSe thin films on topological insulator $(Bi,Sb)_2Te_3$. Direct evidence is obtained for intercalation of MnSe atomic layer into a few quintuple layers of $(Bi,Sb)_2Te_3$, forming either a double magnetic septuple layer (SL) or an isolated single SL at the interface, where one SL denotes a van der Waals building block consisting of B-A-B-Mn-B-A-B (A=$Bi_{1-x}Sb_x$, B= $Te_{1-y}Se_y$). The two types of interfaces (namely TI/mono-SL and TI/bi-SL) have different MPE, which is manifested as distinctively different transport behaviors. Specifically, the mono-SL induces a spin-flip transition with a sharp change at small magnetic field in the anomalous Hall effect of TI layers, while the bi-SL induces a spin-flop transition with a slow change at large field. Our work demonstrates a useful platform to realize the full potential of the magnetic extension approach for pursuing novel topological physics and related device applications.

The interplay between magnetism and topological states can produce a lot of exotic physics, such as the quantum anomalous Hall effect (QAHE) [1],[2], topological magneto-electric (TME) effect [3], and magnetic monopole [4]. To date, two methods have been mainly used to create magnetic order in a topological insulator (TI). One is magnetic impurity doping, which has allowed for the observation of the QAHE [2],[5]. However, the inhomogeneous distribution of magnetic dopants introduces strong disorder [6] and results in extremely low quantization temperature. Another method is to make use of the magnetic proximity effect (MPE) in a heterostructure comprising of a TI and a magnetic insulator (MI) [7],[8]. Many types of MIs have been reported to fabricate the heterostructure, such as EuS [9]-13[13], $Y_3Fe_5O_{12}$ [14]-19[19], $Tm_3Fe_5O_{12}$ [20] , $BaFe_{12}O_{19}$ [21], $Cr_2Ge_2Te_6$ [22], and (Zn,Cr)Te [23]. It is worth noting that the QAHE was reported in (Zn,Cr)Te/$(Bi,Sb)_2Te_3$/(Zn,Cr)Te sandwich structure recently [23]. This is the first observation of the QAHE in a proximity coupled system. But with the Cr dopants induced disorder in (Zn,Cr)Te, the temperature of quantization is below 100 mK, comparable to the case of Cr-doped $(Bi,Sb)_2Te_3$ system. To avoid the inhomogeneity of ferromagnetism, the MI should be an undoped or intrinsic magnetic material. However, no QAHE has been observed so far in the undoped MI/TI systems, which is mainly attributed to the poor interface crystallinity and a weak interfacial coupling. Even with a perfect interface, the MPE could be still smaller than expected, because the lattice mismatching of TI and MI would create a strong interfacial potential [24]. This potential pushes away the Dirac surface states from the interface and hence significantly reduces the size of magnetic gap. Besides, it also introduces trivial interface states in the magnetic gap [24].

An alternative approach, named "magnetic extension", was proposed by M. Otrokov et al. recently [24],[25] to avoid the problems encountered in the above two methods. The key element of this approach is to cover a TI with an ultrathin MI film both structurally and compositionally compatible with the TI. For instance, in $MnSe/Bi_2Se_3$ heterostructures, a MnSe layer can be intercalated into the topmost $Bi_2Se_3$ quintuple layer (QL, stacked as Se-Bi-Se-Bi-Se), and form a $MnBi_2Se_4$ septuple layer (SL, stacked as Se-Bi-Se-Mn-Se-Bi-Se) [26] Due to the perfect matching of the TI and MI layers, the Dirac surface states could penetrate into the MI side and leads to a strong magnetic exchange interaction. A magnetic gap of 90 meV was found in a $MnSe/Bi_2Se_3$ heterostructure by angle-resolved photoemission spectroscopy (ARPES) measurements [26],[27]. In contrast, the MPE in $MnSe/Bi_2Se_3$ heterostructures without such magnetic extension (i.e. without the formation of $MnBi_2Se_4$ at the interface) can only induce a much smaller gap (~8.5 meV predicted by calculations) [24]. Moreover, the magnetic extension approach has advantages including: Fermi level located inside the magnetic gap, no randomly distributed magnetic dopants, and absence of trivial interface states [24]. All these properties make the magnetic extension system particularly attractive for the realization of high temperature QAHE [28], and other exotic physics such as chiral Majorana zero modes [29] that requires extremely low disorder. However, very limited experimental efforts have been made along this direction. In Mogi et al.'s work on $Cr_2Ge_2Te_6/(Bi,Sb)_2Te_3/Cr_2Ge_2Te_6$ [22], the authors suppose the penetration of the TI surface state wave function into $Cr_2Ge_2Te_6$ induces a large anomalous Hall (AH) conductivity despite small magnetization. But the magnetic extension effect in $Cr_2Ge_2Te_6$/TI lacks the support of theoretical calculation and direct experimental evidence. To the best of our knowledge, no electron transport measurement has been reported on the TI/MI heterostructures with clear evidence for the magnetic extension.

In this paper, we investigate the magnetic extension effect in the $MnSe/(Bi,Sb)_2Te_3$ heterostructures. We show clear evidence of self-organized $(Bi,Sb)_2Mn(Te,Se)_4$ SLs forming at the interface (due to the intercalation of MnSe layer into the TI layers) by high-resolution scanning transmission electron microscopy (HR-STEM). Two types of interfaces, $(Bi,Sb)_2Te_3$/mono-SL and $(Bi,Sb)_2Te_3$/bi-SL, are identified. Thickness dependent transport measurements suggest that the magnetic layers in this hybrid system are sufficiently insulating, allowing the electron transport being dominated by the TI surface. The two interfaces show quite different characters in the AH effect due to different magnetic properties. The mono-SL induces a spin-flip feature at small magnetic field, while the bi-SL induces a spin-flop feature at high magnetic field.

MnSe/(Bi,Sb)$_2$Te$_3$ heterostructure thin films were grown on SrTiO$_3$ (111) (STO) substrates by molecule beam epitaxy (MBE) in a UHV system (base pressure 5×10$^{-10}$ mbar). The growth process was monitored with an *in situ* reflection high-energy electron diffraction (RHEED). Most heterostructure samples studied in this work are composed of a 5 nm thick MnSe thin film and a 13nm thick (Bi,Sb)$_2$Te$_3$ thin film. Samples were patterned into 200 μm wide Hall bars devices (schematically depicted in Fig. 1(a)), and the transport measurements were carried out in a helium vapor flow cryostat (1.6 K). STO substrate with high dielectric constant is utilized as back gate to tune the chemical potential of TIs and realize carrier density variation of ∼ 3×10$^{13}$ cm$^{-2}$ with back-gate voltage ($V_g$) between ± 210 V for typical substrate thickness of 0.5mm. Atomic force microscopy (AFM), X-ray diffraction (XRD), and HR-STEM have been performed for structure characterization. Fig. 1(b) shows a typical cross-section TEM image of the MnSe/(Bi,Sb)$_2$Te$_3$/STO (5nm/13nm) sample measured along [$\bar{1}$100]. At the interface, one can clearly see the formation of (Bi,Sb)$_2$Mn(Te,Se)$_4$ SLs, consisting of seven monoatomic layers with stacking sequence Te(Se)-Bi(Sb)-Te(Se)-Mn-Te(Se)-Bi(Sb)-Te(Se). The (Bi,Sb)$_2$Mn(Te,Se)$_4$ SL is presumably isostructural to the previously studied MnBi$_2$Se$_4$ and MnBi$_2$Te$_4$ [-30-(38)]. In the image, the Mn intercalated layers are arranged as SL-QL-SL-SL-QL along c axis. We note that the layer sequence of SL and QL are not uniform along the plane of sample. Alternatively speaking, the first two layers nearby (Bi,Sb)$_2$Te$_3$ are distributed as either SL-QL (mono-SL) or SL-SL (bi-SL) in different interfacial regions. More representative TEM images are provided in Figure S1 of the Supporting Information, showing the observation of mono-SL and bi-SL in contact with (Bi,Sb)$_2$Te$_3$ in the same sample.

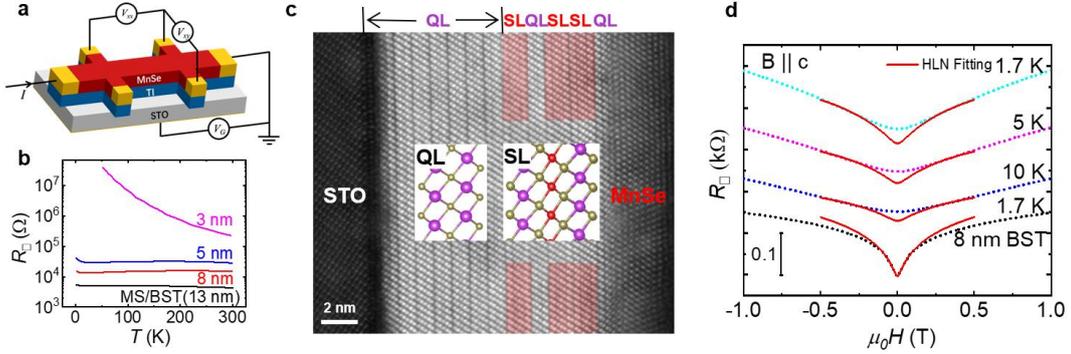

**Figure 1.** (a) Schematic sketch of a back-gated Hall bar device fabricated on the MnSe/(Bi,Sb)$_2$Te$_3$ heterostructure. (b) Temperature dependences of sheet resistance R$_□$ of heterostructures with different (Bi,Sb)$_2$Te$_3$ growth thickness of 3, 5, 8 and 13 nm, respectively. (c) Cross-section scanning transmission electron microscopy (STEM) image of the nominal MnSe/(Bi,Sb)$_2$Te$_3$(5nm/13nm) grown on the STO substrate. The insets are the sketches of quintuple layer (QL) and septuple layer (SL, due to intercalation of Mn to QL). QL consists of Te-Bi(Sb)-Te-Bi(Sb)-Te and SL consists of Te(Se)-Bi(Sb)-Te(Se)-Mn-Te(Se)-Bi(Sb)-Te(Se) monoatomic layers. Te(Se): yellow solid circle; Bi(Sb): purple solid circle; Mn: red solid circle. A mixed interfacial layers of SL-QL-SL-SL-QL is observed along c axis (red shaded area). (d) Magnetoresistances of the heterostructure shown in (c) measured at different temperatures from 1.7 K to 10 K, and 8 nm thick (Bi,Sb)$_2$Te$_3$ thin film at 1.7 K. Hikami-Larkin-Nagaoka (HLN) equation fitting curves are shown in the red lines. The (Bi,Sb)$_2$Te$_3$ thin film can be well fitted to HLN equation that describes the weak-antilocalization effect (WAL) behavior at low field, while MnSe/(Bi,Sb)$_2$Te$_3$ deviates strongly from HLN fitting.

Fig. 1(b) displays the temperature dependent sheet resistance $R_\square$ of samples with different (Bi,Sb)$_2$Te$_3$ growth thickness (3~13 nm). Due to the formation of Mn intercalated layers, the actual thickness of TI is reduced. The $R_\square$ of samples is measured to be larger than that of (Bi,Sb)$_2$Te$_3$ film with the same nominal thickness. For example, MnSe/(Bi,Sb)$_2$Te$_3$ (5nm/13nm) is actually "MnSe/Mn intercalated layers/(Bi,Sb)$_2$Te$_3$" (5nm/5nm/8nm), and its sheet resistance is found to be close to that of 8nm (Bi,Sb)$_2$Te$_3$ but larger than that of 13nm (Bi,Sb)$_2$Te$_3$ (A detailed analysis is provided in part two of the Supporting Information). For MnSe/(Bi,Sb)$_2$Te$_3$ (5 nm/3 nm), $R_\square$ is about 50 MΩ at 50 K. Thus it strongly indicates that the magnetic layers composed of the intercalated layers and the MnSe layers are electrical insulating, in stark contrast to the metallic MnBi$_2$Te$_4$ [35-40]. Electron transport in the heterostructure is therefore mainly dominated by the (Bi,Sb)$_2$Te$_3$ layer, without the detrimental shunting from the magnetic layer. This is further confirmed by the longitudinal magnetoresistance (MR) data recorded in perpendicular magnetic fields. As shown in Fig. 1(d), the MR is positive, instead of the negative MR observed in MnBi$_2$Te$_4$ layers [37],[38].

The MR of MnSe/(Bi,Sb)$_2$Te$_3$ heterostructure is nearly parabolic at low magnetic fields, differing strikingly from the cusp-like MR seen in TIs grown on nonmagnetic substrates [41]. As illustrated in Fig. 1(d), (Bi,Sb)$_2$Te$_3$/STO exhibits the typical weak-antilocalization effect (WAL) behavior and can be fitted well to the Hikami-Larkin-Nagaoka (HLN) equation at low field. In contrast, the low field MR of MnSe/(Bi,Sb)$_2$Te$_3$ deviates from HLN fitting. The strong suppression of WAL in MnSe/(Bi,Sb)$_2$Te$_3$ provides a signature for the broken time-reversal symmetry (TRS) due to the interfacial magnetic interaction.

Hall measurements provide further evidence for the interfacial magnetic exchange interaction in MnSe/(Bi,Sb)$_2$Te$_3$. Fig. 2 displays the magnetic field dependent Hall resistance $R_{xy}$ at varied back gate voltage $V_G$ from −210 V to 210 V. The AH effect with prominent hysteresis loop is clearly visible. The slope of the background at high field evolves from negative to positive with increasing $V_G$, indicating the change of the dominant carriers from p- to n-type. The coexistence of two type carriers leads to a nonlinear feature of the background, especially around the charge neutral point at -160 V where the hysteresis loop is most obvious. Away from the neutral point, the ordinary Hall background dominates and largely overwhelms the AH component.

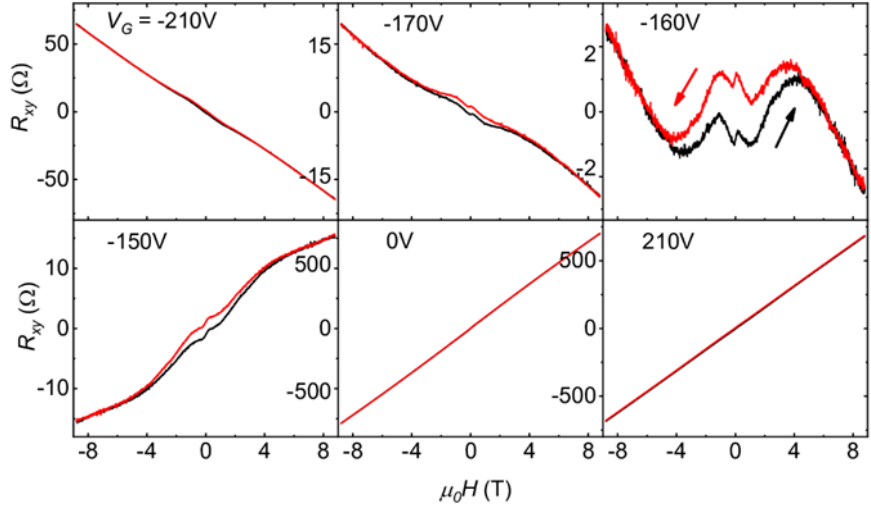

**Figure 2.** Gate-voltage tuning of Hall resistance $R_{xy}$ in the nominal MnSe/(Bi,Sb)$_2$Te$_3$(5nm/13nm) heterostructure at T =1.7 K. The gate voltage is varied from −210 V to 210 V with the carriers changed from p-type (negative slope) to n-type (positive slope). The red and black curves represent magnetic field sweeping to the negative and positive directions, respectively.

Fig. 3(a) displays the temperature dependences of Hall resistances, which provide evidence for the interfacial magnetic interaction. As the temperature is increased, the hysteresis loop gets smaller gradually. The temperature at which hysteresis almost disappears (denoted here as $T_t$, magnetic phase transition temperature) is estimated to be 8 K according to the Hall data, consistent with the results in Fig. 3(b) where the hysteresis loop of parallel field MR also disappears around 8 K. As the measured $T_t$ is much closer to the Neel temperature ($T_N$) of MnBi$_2$Se$_4$ about 14 K [42] than that of MnSe about 197 K [43], the magnetism of (Bi,Sb)$_2$Te$_3$ should be induced by the exchange interaction with the interfacial Mn intercalated layers rather than with the upper MnSe layers.

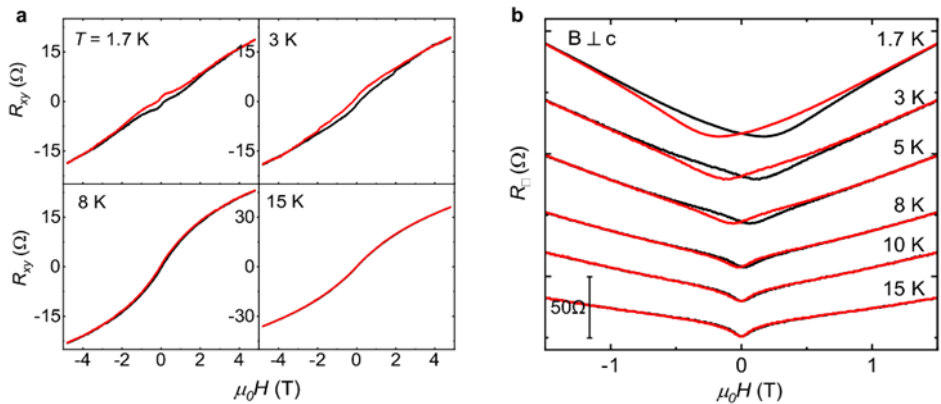

**Figure 3.** Hall resistance $R_{xy}$ (a) and in-plane sheet resistance $R_\square$ (b) plotted as a function of magnetic field from 1.7 K to 15 K in MnSe/(Bi,Sb)$_2$Te$_3$. With increasing temperature, the hysteresis loop becomes smaller and almost disappears above 8 K. The scale bar is 50 Ω.

The parallel field MR in Fig. 3(b) shows a hysteresis with the minimum value near

$H_c$ ($H_c$ is the coercive field of AH resistance in the parallel field. The corresponding data is shown in the Supporting Information). Similar phenomenon has been observed in impurity-doped $Bi_2Te_3$[44] and $Bi_2Se_3$/EuS [9]. The enhanced conductivity at the magnetization reversal is induced by the chiral conducting modes propagating along the domain walls that form between the domains with canted magnetization at the interface. According to our previous results on $Bi_2Se_3/BaFe_{12}O_{19}$, parallel field MR is more sensitive to magnetism than perpendicular field MR [21]. This applies here and explains why the hysteresis is observed in the parallel field MR (Fig. 3(b)) but not in the perpendicular field MR (Fig. 1(d)). We need to note here that, a square hysteresis loop is observed for the AH resistance in the parallel field (Figure S3 in the Supporting Information). As the AH effect is typically sensitive to the perpendicular magnetization, the square-shaped hysteresis should be contributed by the canted or more complicated spin structures due to the defects at the interface.

As $R_{xy} = R_{AH} + R_{OH}$, the AH resistances ($R_{AH}$) can be obtained by subtracting the ordinary component ($R_{OH}$) from the total Hall resistances. Due to the influence of multi-carriers, the ordinary Hall resistances in most cases are not strictly linear with the magnetic field for the entire gate tuning range. This complicates the separation of the AH signal from the ordinary Hall resistance. In this work, we limit our quantitative analysis to the gate voltages with weak nonlinearity in the ordinary Hall background [Supporting Information]. The obtained AH resistance is converted to the AH conductivity by $\sigma_{AH} = R_{AH}/(R_{xy}^2 + R_{xx}^2)$. Fig. 4(a) shows the derived $\sigma_{AH}$ curve which changes very rapidly around zero field but changes much slower at high fields. This is due to the different magnetic interactions at the two types of interfaces (TI/mono-SL and TI/bi-SL) formation in our samples, as analyzed in detail below.

The magnetic properties of bi-SL should be similar to that of thin flake $MnBi_2Te_4$ with thickness down to a few SLs [30],[33],[37]-[38], which shows an external magnetic field driven spin-flop transition with a small hysteresis loop of AHE at low field and magnetization saturation at high field above 5 T [30],[37],[38]. As for the properties of mono-SL, the recent exploration of the bulk $MnBi_{2n}Te_{3n+1}$ (n = 1, 2 and 3) ternary system can offer some helpful reference [-45-47]. The original compound $MnBi_2Te_4$ is identified as interlayer antiferromagnet, with the neighboring ferromagnetic (FM) SLs coupled antiparallel to each other and the easy axis of staggered magnetization pointing perpendicular to the layers. In comparison, $MnBi_{2n}Te_{3n+1}$ consists of alternating [$MnBi_2Te_4$] and n[$Bi_2Te_3$] layers, and has much weaker interlayer exchange interaction than that of $MnBi_2Te_4$ by increasing the interlayer distance with extra insulating [$Bi_2Te_3$] spacer layers in between the SL blocks [45],[46]. Consequently, $MnBi_4Te_7$ (similar case of mono-SL) reveals a weaker interlayer exchange interaction that leads to a lower $T_N$ of 12K than $MnBi_2Te_4$ (25K). Unlike the spin-flop transition in $MnBi_2Te_4$, a spin-flip transition takes place in bulk $MnBi_4Te_7$ at a much lower field of 0.15 T and saturates at 0.22 T, more than one order of magnitude lower than that of $MnBi_2Te_4$ [45].

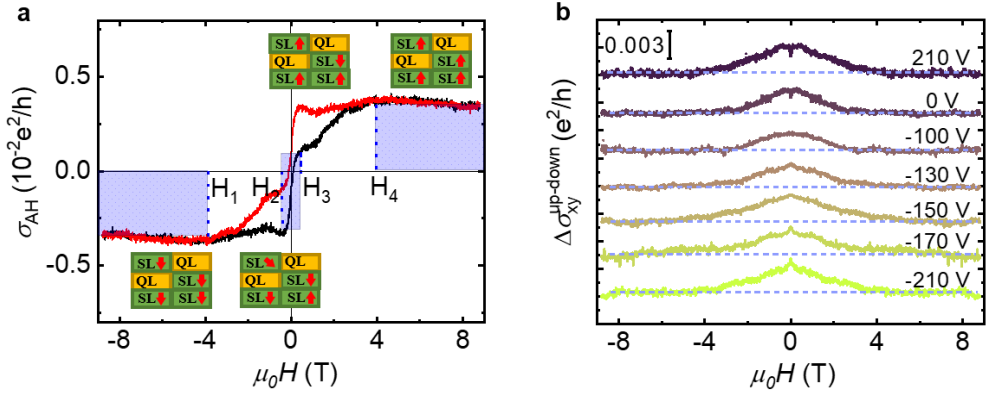

**Figure 4.** (a) Magnetic field dependence of anomalous Hall conductance $\sigma_{AH}$ at $V_G = 0$ V. Schematic drawing of the spin switching process in the Mn intercalated layers, corresponding to the magnetic field (dashed line) of $H_1$ to $H_4$. (b) Magnetic field dependence of Hall conductance $\Delta\sigma_{xy}^{up-down}$ (the difference value of sweep-up and sweep-down Hall conductance) at different gate voltages. All the Hall data are taken from the heterostructure sample at 1.7 K. The scale bar is 0.003 $e^2$/h.

In MnSe/(Bi,Sb)$_2$Te$_3$, the TI layers exhibit characteristics of both mono-SL and bi-SL interfacial layers, as demonstrated in Fig. 4(a), where both a spin-flip transition and a spin-flop transition are observed. The AH conductivity undergoes a rapid change within the low field range from $H_2$ to $H_3$. In this spin-flip process, the magnetization of mono-SL suddenly reverses to the same direction with a small saturation field about 0.2 T, close to that of MnBi$_4$Te$_7$. Within the high field range ($H_1$~$H_2$ and $H_3$~$H_4$), the spin-flop transition occurs in bi-SL, with a much slower change of AH conductivity. In the spin-flop process, the system first transits from an AFM to a metastable canted AFM state, where the magnetization is gradually polarized until the magnetic moment is completely saturated and the whole system arrives in a FM state ultimately. In addition, we performed control experiments on Mn doped (Bi,Sb)$_2$Te$_3$ and MnSe/(Bi,Sb)$_2$Te$_3$ heterostructure without Mn intercalation (part five and part six in the supporting information). These samples show quite different transport behavior, further confirming the above complicated AH effect are induced by the magnetic extension of the Mn intercalated layers.

To avoid the complications arising from the nonlinear ordinary Hall signals, we directly subtract the sweep-down curve from the sweep-up curve. The resulted difference in the Hall conductivity, $\Delta\sigma_{xy}^{up-down}$ is shown in Fig. 4(b). It reflects the response of AH effect to the interfacial magnetic interaction below the magnetic saturation field (about 4 T). The overall shape remains about the same at different voltages. The gate-voltage tuning has very limited influence on the AH effect, despite that the ordinary Hall resistance is strongly influenced [Fig. 2]. This can be attributed to the weak influence of the back-gate on the top surface states. According to previous studies, electrostatic screening of the bottom surface as well as the bulk states make it hard to tune the chemical potential on the top surface with a back gate unless the film is very thin [48],[49]. The insensitivity of the AH effect to the gate voltage suggests the observed AH conductivities originate from the interfacial magnetic coupling, rather than the unwanted doping effects in the bulk. Besides, we note a small peak of $\Delta\sigma_{xy}^{up-down}$ around zero field where the spin flip takes place. As the dominating carriers change from n-type to p-type with decreasing $V_G$, the peak transits from downward to upward gradually around zero field. This phenomenon may be induced by the sensitivity of magnetic exchange interaction to the carrier type during the process of rapid spin flip. But further research is needed to clarify it.

It is noteworthy that the transport properties of MnSe/(Bi,Sb)$_2$Te$_3$ heterostructures observed in this work are quite different from of MnBi$_2$Te$_4$ or MnBi$_4$Te$_7$ [37],[38],[45],[46] even though both can be attributed to the broken TRS due to the magnetic interaction. In perpendicular magnetic fields, the former is positive (Fig. 1(d)), while the latter is negative. The former is dominated by the TI surface states under influence of interfacial coupling. Thus, it does not deviate strongly from the WAL behavior if the magnetic field is not too weak. In the latter, the negative MR is likely related to spin dependent scatterings related to unaligned Mn spins [45],[46].

The magnetic extension system possesses unique advantages over the other systems of coupling magnetism and topology. By growing MnSe on a selected region of TI, one can break the TRS locally, which cannot be achieved in MnBi$_2$Te$_4$. This area-selective gap opening is very critical for the investigation of many emergent phenomena including the half-integer QAHE [50] and the Majorana fermion edge states [51],[52]. Due to the influence of bulk defects, the AH effect of our present sample is still small (~10 Ω). But much larger AH effect could be expected in this system by optimizing the ratio of Bi/Sb and reducing the bulk conduction. Besides, the growth temperature of MnSe and TI is close, which facilitates the fabrication of MnSe/TI/MnSe structure. With both the top and bottom surface state of TI magnetized by MnSe through the magnetic extension effect, it's hopeful to enhance the temperature of QAHE.

In summary, we have carried out a systematic study on MnSe/(Bi,Sb)$_2$Te$_3$ heterostructures. We provide direct proof of the existence of Mn intercalated SLs at the interface by HR-TEM. Transport experiments show a strong exchange interaction between the SLs and the surface state of (Bi,Sb)$_2$Te$_3$ as a result of magnetic extension effect, as verified by the suppressed WAL in MR and the hysteresis in AHE. More interestingly, we observe the coexistence of spin-flip and spin-flop transitions originating respectively from the mono-SL and bi-SL interfacial structures. Our results show that MnSe/TI heterostructure could serve as an ideal system for exploring QAHE, quantized TME effect, and many other exotic topological quantum effects.

■ AUTHOR INFORMATION
**Corresponding Authors**
*Jing Teng. E-mail: jteng@iphy.ac.cn
*Qinghua Zhang. E-mail: zqh@iphy.ac.cn
**Notes**
The authors declare no competing financial interest.

■ ACKNOWLEDGEMENTS:
This work was supported by the National Key Research and Development Program (Project No. 2016YFA0300600), the National Science Foundation of China (Project No. 61425015, and No. 11604374), the National Basic Research Program of China (Project No. 2015CB921102), and the Strategic Priority Research Program of Chinese Academy of Sciences (Project No. XDB28000000).

# Supporting Information:

1. **Two kinds of interfacial structures, SL-QL (mono-SL) and SL-SL (bi-SL) in the heterostructure sample**

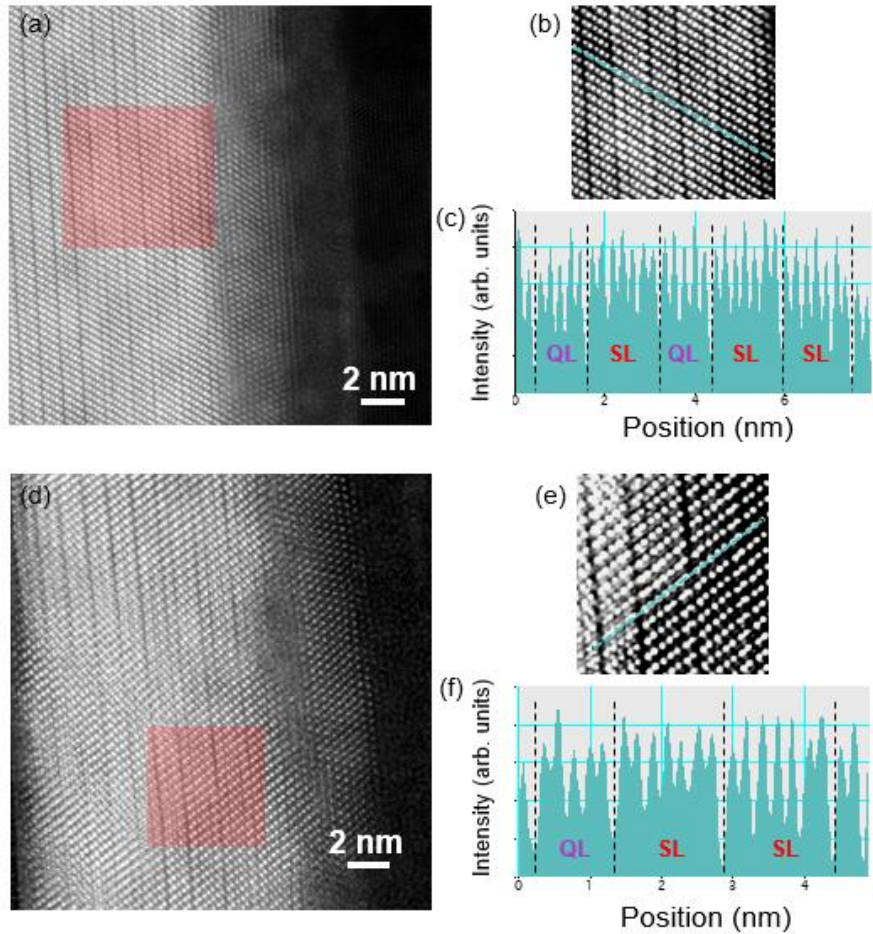

Figure S1: (a, d) Representative cross-section scanning transmission electron microscopy (STEM) images of MnSe/(Bi,Sb)$_2$Te$_3$ (5nm/13nm) grown on the STO substrate. (b, e) Zoom-in views of (a, d). (c, f) Intensity distribution of STEM along the cut lines (light blue) in (b, e).

2. **The insulating nature of Mn intercalated layers**

Figure S2 shows the sheet resistance R$_\square$ of MnSe/(Bi,Sb)$_2$Te$_3$ heterostructure samples with different (Bi,Sb)$_2$Te$_3$ growth thickness (3~13 nm) and R$_\square$ of pure (Bi,Sb)$_2$Te$_3$ (and Bi$_2$Se$_3$) thin films (adapted from PRL 114, 216601 (2015)). R$_\square$ of pure

TI with thickness $\geq 8$ nm is around 3 k$\Omega$, similar to R$_\square$ of MnSe/(Bi,Sb)$_2$Te$_3$ heterostructure with 13 nm (Bi,Sb)$_2$Te$_3$. R$_\square$ of pure (Bi,Sb)$_2$Te$_3$ of 5 nm is 10 k$\Omega$, similar to that of MnSe/(Bi,Sb)$_2$Te$_3$ heterostructure with 8 nm (Bi,Sb)$_2$Te$_3$. R$_\square$ of pure (Bi,Sb)$_2$Te$_3$(Bi$_2$Se$_3$) of 3 nm is about tens of k$\Omega$, similar to that of MnSe/(Bi,Sb)$_2$Te$_3$ heterostructure with 5 nm (Bi,Sb)$_2$Te$_3$. Thereforefore, it strongly indicates that the real thickness of (Bi,Sb)$_2$Te$_3$ is 2~5 nm thinner than the nominal one in the heterostructure, consistent with the TEM observation that shows 2~5 quintuple layers have transformed to septuple layers. The Mn intercalated layers (less than 3 SL in the MS/BST samples) show an insulating character with the (Bi,Sb)$_2$Te$_3$ layers dominating the transport.

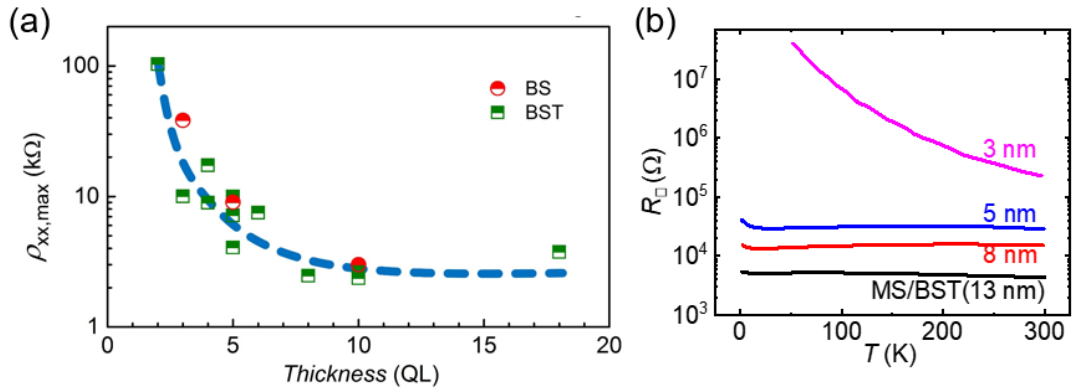

Figure S2. (a) Adapted from Phys. Rev. Lett. 114, 216601 (2015), $\rho_{xx,\max}$ versus film thickness of (Bi,Sb)$_2$Te$_3$ and Bi$_2$Se$_3$ samples. The dashed line shows the qualitative tendency of the thickness dependence. (b) Temperature dependences of sheet resistance R$_\square$ of MnSe/(Bi,Sb)$_2$Te$_3$ with different (Bi,Sb)$_2$Te$_3$ growth thickness of 3, 5, 8 and 13 nm, respectively.

## 3. Gate-voltage tuning in the plane field

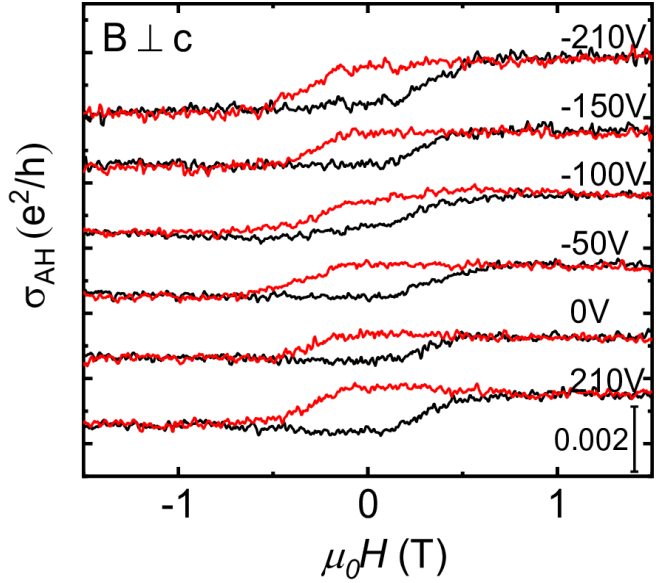

Figure S3: In-plane magnetic field dependences of anomalous Hall conductance $\sigma_{AH}$ at different gate voltages. In the voltage range from -210 V to 210 V, the magnitude of $\sigma_{AH}$ is almost unchanged with a hysteresis loop. Data are taken at 1.7 K and the scale bar is 0.002 $e^2/h$.

## 4. Extraction of anomalous Hall resistance

The ordinary Hall resistances ($R_{OH}$) are not strictly linear with respect to the magnetic field for the entire gate voltage range due to the influence of multi-carriers in the sample. Thus we cannot extract the AH resistances ($R_{AH}$) from the total Hall resistances by simply subtracting the linear background. For slightly nonlinear $R_{OH}$, a polynomial fitting can be applied. Here, we take the case of $V_G = 0$ V as an example. The following is the detailed fitting process as shown in Supplementary Figure 2.

As $R_{OH}$ is an odd function, the equation of $R_{xy} = aB^5 + bB^3 + cB + d$ is used, where $a$, $b$, $c$, $d$ are coefficients, and $d$ is an offset to remove the influence of $R_{AH}$. We first fit $R_{xy}/B$ with the fitting range from -8.8 to -5 T, where there is no contribution of $R_{AH}$. And the parameters a, b, c, and d are determined (panel (b)). Then the ordinary Hall resistance: $R_{OH} = aB^5 + bB^3 + cB$ is obtained as shown in panel (c). Finally, $R_{AH}$ is extracted by subtracting $R_{OH}$ from the total Hall resistance with $R_{AH} = R_{xy} - R_{OH}$ (panel (d)).

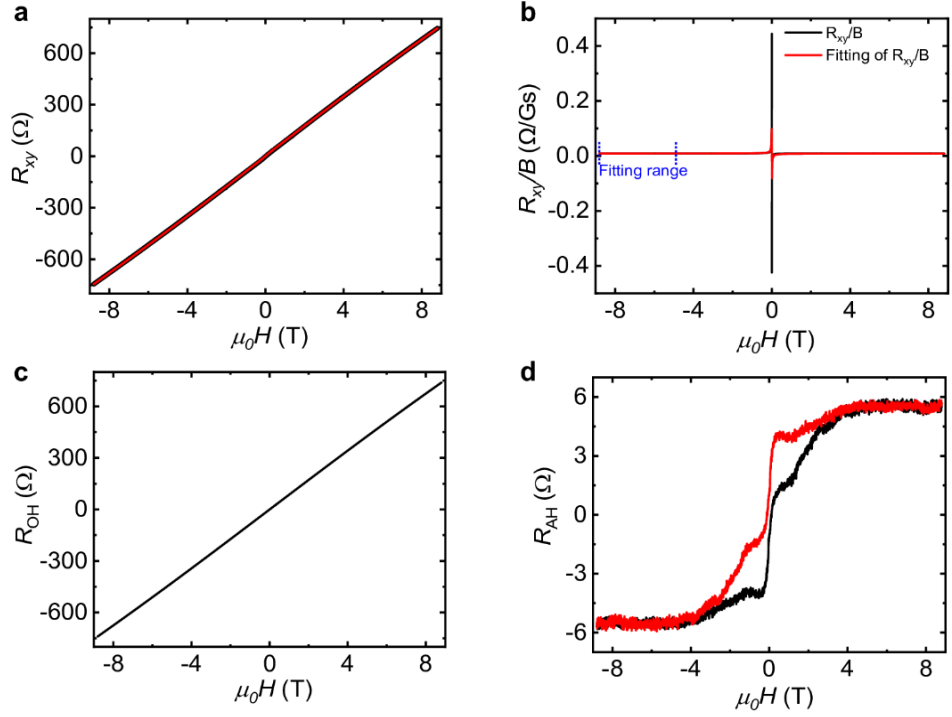

Figure S4: The extraction process of anomalous Hall resistance ($R_{AH}$) at $V_G = 0$ V and $T = 1.7$ K. (a) Magnetic field dependence of Hall resistance ($R_{xy}$). (b) Magnetic field dependence of $R_{xy}/B$ (black line) and a transformed polynomial fit (red line), i.e. $R_{xy} = aB^4 + bB^2 + c + d/B$. The fitting range is from -8.8 to -5 T. (c) The ordinary Hall resistance ($R_{OH}$), obtained by using parameters a, b and c extracted from the above fitting. (d) $R_{AH}$ extracted by subtracting $R_{OH}$ from $R_{xy}$, i.e. $R_{AH} = R_{xy} - R_{OH}$.

## 5. The control experiments on MnSe/(Bi,Sb)$_2$Te$_3$ heterostructure without Mn intercalated layers

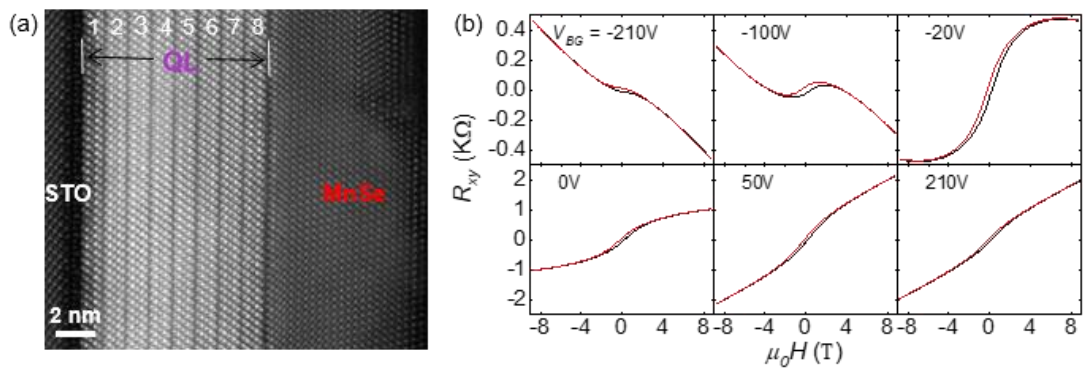

Figure S5: Structure and transport results of MnSe/(Bi,Sb)$_2$Te$_3$ without any Mn intercalated layer at the interface. The growth temperature of MnSe on (Bi,Sb)$_2$Te$_3$ is controlled to be lower than 350 ºC to avoid the intercalation of Mn atoms. (a) Cross-section scanning transmission electron microscopy (STEM) image of the heterostructure grown on the STO substrate. (b) Gate-voltage tuning of Hall resistance $R_{xy}$ at T =1.7 K. The gate voltage is varied from −210 V to 210 V with the carriers changed from p-type (negative slope) to n-type (positive slope).

## 6. The control experiments on 13 nm Mn-doped (Bi,Sb)$_2$Te$_3$ thin film

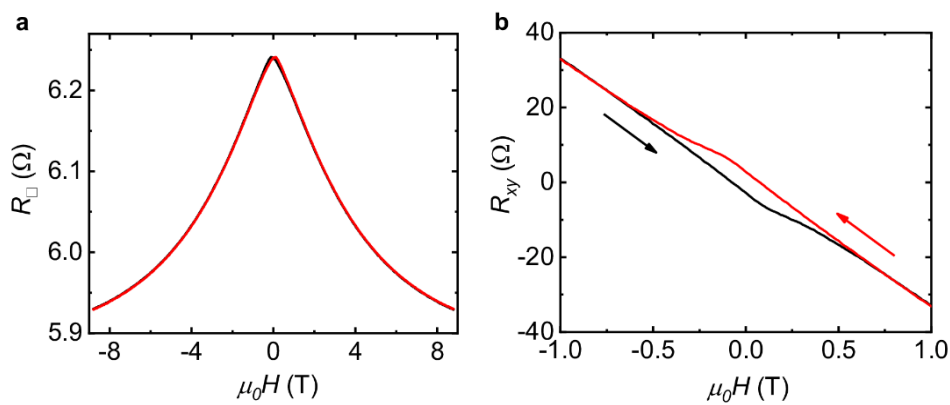

Figure S6: Out-of-plane magnetic field dependences of sheet resistance $R_\square$ (a) and Hall

resistance $R_{xy}$ (b) of 13 nm Mn-doped $(Bi,Sb)_2Te_3$ thin film at 3 K. The magnetoresistance is completely negative and the hysteresis shape of $R_{xy}$ in low field displays a regular rectangle, in sharp contrast to the behaviors of MnSe/$(Bi,Sb)_2Te_3$ heterostructure (See Figure 1(d) and Figure 2 in the main text). This control experiment indicates that the magnetic order in MnSe/$(Bi,Sb)_2Te_3$ heterostructure does not originate from the diffusion of Mn atoms to the interface.